\documentclass[twocolumn,showpacs,superscriptaddress]{revtex4}
\usepackage{amsmath}
\usepackage{graphicx}
\usepackage{dcolumn}
\usepackage{bm}

\begin{document}
\title{Teleportation on a quantum dot array}
\author{F. de Pasquale}
\email{ferdinando.depasquale@roma1.infn.it}
\author{G. Giorgi}
\affiliation{INFM Center for Statistical Mechanics and Complexity}
\affiliation{Dipartimento di Fisica, Universit\`{a} di Roma La
Sapienza, P. A. Moro 2, 00185 Rome, Italy}
\author{S. Paganelli}
\affiliation{Dipartimento di Fisica, Universit\`{a} di Roma La
Sapienza, P. A. Moro 2, 00185 Rome, Italy}
\affiliation{Dipartimento di Fisica, Universit\`{a} di Bologna,
Via Irnerio 46, I-40126, Bologna, Italy}
\date{\today}
\begin{abstract}
We present a model of quantum teleportation protocol based on a
double quantum dot array. The unknown qubit is encoded using a
pair of quantum dots, with one excess electron, coupled by
tunneling, . It is shown how to create a maximally entangled state
using an adiabatically increasing\ Coulomb repulsion between
different dot pairs. This entangled state is exploited to perform
teleportation again using an adiabatic coupling between itself and
the incoming unknown state. Finally, a sudden separation of Bob's
qubit allows a time evolution of Alice's which amounts to a
modified version of standard Bell measurement. A transmission over
a long distance could be obtained by considering the entangled
state of a chain of coupled double quantum dots.
\end{abstract}
\maketitle

\section{Introduction}

Building a quantum computer is a challenge of modern physics
\cite{bendiv}.\ Required ingredients are entanglement, quantum
channels with high fidelity for information transfer, one and two
(unconditional and conditional) qubit gates. During last years
many efforts have been done in different fields to design a
feasible quantum computer. A very efficient protocol to transfer a
state from one place to another is represented by quantum
teleportation \cite {tele}.

Experimental realizations have been performed with optical systems
\cite {optics} and NMR techniques \cite{NMR} with spatially
localized qubits since the degree of freedom of codification were
represented by polarization for photons and spin for molecules.

In a solid state approach teleportation schemes have been proposed
using spin \cite{feinberg} and vacuum- single exciton qubit
\cite{reina}\ on semiconducting quantum dots. The first one of
these schemes considers the use of a circuit of three normal and
two superconducting QDs; the teleportation is achieved by
measuring the spin-polarized current through the dot array, and
the maximum distance of information transport limited by the
coherence length of the superconducting device.

Ground state-single exciton qubits in quantum dots have been also
proposed for quantum computation architecture using non adiabatic
processes \cite {zanardi}. Recently the realization of number
state qubits had interesting developments. In this framework a
teleportation protocol has been successfully implemented with
photons \cite{lombardi}. Here we propose a solid state
implementation based on a double quantum dot (DQD) array that
permits to create entanglement and to perform quantum
teleportation. Our qubit is represented by a pair of spatially
separated quantum dots coupled by tunneling. Interactions between
different qubits are driven by means of Coulomb repulsion. These
charge (or number) states are characterized by one excess electron
occupying the first or the second dot, creating an isolated\ well
defined quantum two level system.

A similar configuration \cite{mompart} can be adopted in an ion
trap quantum computing device localizing a neutral atom in one
member of a pair of trapping potentials. Despite decoherence
obstacles, principally due to cotunneling effects, electron-phonon
interactions and background charge fluctuations, coherent
oscillations in DQD systems have been recently experimentally
observed \cite{fujisawa} and therefore it becomes more and more
interesting\ to analyze the potentialities of these schemes. In
order to obtain the desired operations, both adiabatic and
nonadiabatic changes in the interaction parameters are requested.
The charge transport through dots using adiabatic variations of
pairing parameters has been extensively described \cite{brandes},
while in a different scenario (NMR) quantum adiabatic gates have
been experimentally achieved \cite{steffen}.

We preliminary show how to create Alice's general unknown state
starting from an initial one by tunneling. Next we show how to
generate the entangled state of two qubits, which represents the
support for teleportation by adiabatic switching on Coulomb
interaction. The unknown state is then entangled with the support
state again by adiabatic switching on of Coulomb interaction.
Bob's qubits is obtained by detachment of last pair. From now on
the system evolves for a suitable time at which the measurement of
a number state of one of Alice's dots ends the teleportation
process. By a classical communication of Alice's measurement to
Bob the unknown state can be reconstructed. This simple scheme can
be easily extended to an $N$ pairs teleportation support and is
presumably robust with respect to electron-phonon interaction.
Decoherence effects will be mainly due to coupling with external
leads \cite{fujisawa}.
\section{Entangled states}

Let us consider a system composed by a DQD with just one excess
electron with respect to the ground state. This system represents
a charge (or number) qubit with basis elements $\left\vert
0_{1},1_{2}\right\rangle $ and $\left\vert
1_{1},0_{2}\right\rangle $. If dots are coupled by tunneling, in
the presence of a vector potential $\mathbf{A}$ directed from dot
$1$ to dot $2$, the system is described by the following
Hamiltonian:
\begin{equation}
H_{12}=-(we^{-i\varphi }a_{1}^{\dag }a_{2}+we^{i\varphi }a_{2}^{\dag
}a_{1})+\epsilon (a_{2}^{\dag }a_{2}+a_{1}^{\dag }a_{1})
\end{equation}
where $a_{i}(a_{i}^{\dag })$ represents the annihilation
(creation) fermionic operator on the site $i$ and $\varphi
=\frac{eA}{\hbar }$. For sake of simplicity and without loss of
generality we shall assume $\epsilon =0$. This Hamiltonian has
eigenvalues $E^{\pm }=\pm w$ associated to the eigenvectors
$\left\vert E^{\pm }\right\rangle =\frac{1}{\sqrt{2}}\left[ e^{\mp
i\varphi }\left\vert 0_{1},1_{2}\right\rangle \mp \left\vert
1_{1},0_{2}\right\rangle \right] $.

If, assuming $\hbar =1$, we suppose that the system is in a
particular state at $t=0$ (e.g. $\left\vert \phi (0)\right\rangle
=\left\vert 0_{1},1_{2}\right\rangle $), the time evolution
creates a coherent superposition:
\begin{equation}
\left\vert \phi (t)\right\rangle =\cos wt\left\vert 0_{1},1_{2}\right\rangle
+i\sin wte^{2i\varphi }\left\vert 1_{1},0_{2}\right\rangle
\end{equation}
Thus, by instantaneously switching off the tunneling at a suitable time $%
\bar{t}$, we can encode a qubit $\left\vert \phi
(\bar{t})\right\rangle =\left\vert \chi \right\rangle =\alpha
\left\vert 0_{1},1_{2}\right\rangle +\beta \left\vert
1_{1},0_{2}\right\rangle $.

The entangled support for teleportation is an array of four QDs
labelled with subscripts $3,4,5,6$ disposed as indicated in figure
1. The Hamiltonian
\begin{equation}
H_{3456}=-w\left( a_{3}^{\dag }a_{4}+a_{5}^{\dag }a_{6}+h.c.\right) +U\left(
t\right) \left( n_{3}n_{6}+n_{4}n_{5}\right)  \label{Hamiltonian1}
\end{equation}
takes into account both of tunneling interaction along vertical
lines and of Coulomb repulsion along horizontal lines. Here
$n_{i}=a_{i}^{\dag }a_{i}$ is the occupation number operator on
the site $i$. Double occupation on a single dot, as well as double
occupation on a DQD will be completely neglected. Lateral interdot
tunneling interdiction has been considered also
in \cite{loss}. Starting from $U(0)=0$ the Hamiltonian is separable: $%
H_{3456}=H_{34}+H_{56}$. For convenience we shall assume that the
system is prepared in its ground state:

\begin{equation}
\left\vert \psi(0)\right\rangle =\frac{1}{2}\left[ \left\vert
0_{3},1_{4}\right\rangle +\left\vert 1_{3},0_{4}\right\rangle \right] \left[
\left\vert 0_{5},1_{6}\right\rangle +\left\vert 1_{5},0_{6}\right\rangle %
\right]  \label{ground}
\end{equation}

An adiabatic growth of Coulomb repulsion between dot localized on
the same row will create a near maximally entangled state. Here
adiabatic means slow with respect to the lower frequency of the
system. Due to the adiabatic theorem \cite{messiah}, the overall
system will remain in its instantaneous ground state. The
asymptotic behavior is a good approximation of a maximally
entangled state in the limit of $w/U\rightarrow 0$:
\begin{align}
\left\vert \psi (t\rightarrow \infty )\right\rangle & =\frac{1}{\mathcal{N}}%
[\left\vert 0_{3},1_{4},0_{5},1_{6}\right\rangle +\left\vert
1_{3},0_{4},1_{5},0_{6}\right\rangle +  \notag \\
& -\frac{1}{4w}\left( -U+\sqrt{U^{2}+16w^{2}}\right) \left\vert
1_{3},0_{4},0_{5},1_{6}\right\rangle +  \notag \\
& -\frac{1}{4w}\left( -U+\sqrt{U^{2}+16w^{2}}\right) \left\vert
0_{3},1_{4},1_{5},0_{6}\right\rangle ]  \label{tinfinito}
\end{align}
where $\mathcal{N}$ is a normalization coefficient and
$U=U(t\rightarrow \infty )$; in the limit $w/U\rightarrow 0$
$\mathcal{N}\sim 1/\sqrt{2}$. It can be shown that a maximally
entangled state is a good approximation even in the case of an
array with $N>2$ dot pairs. Bearing in mind the limit of
approximation we consider as starting point for the following
manipulation the state
\begin{equation}
\left\vert \psi (t)\right\rangle =\frac{1}{\sqrt{2}}\left[ \left\vert
0_{3},1_{4},0_{5},1_{6}\right\rangle +\left\vert
1_{3},0_{4},1_{5},0_{6}\right\rangle \right]
\end{equation}
The same entangled state can be obtained by quantum annealing procedure in
the presence of a time independent interaction.

\begin{figure}
  \includegraphics[width=2.9075in]{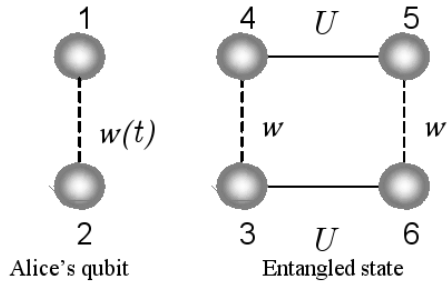}
  \caption{QD's $1$ and $2$ represent the unknown qubit to
teleport. QD's $3\rightarrow 6$ are in the entangled state $\frac{1}{\protect%
\sqrt{2}}\left[ \left\vert 0_{3},1_{4},0_{5},1_{6}\right\rangle
+\left\vert 1_{3},0_{4},1_{5},0_{6}\right\rangle \right] $.
Initially the system are separated. Solid lines represent
tunneling, while dash lines represent Coulomb
repulsion}\label{fig1}
\end{figure}

\section{The teleportation}

The creation of these entangled states enables us to implement a
quantum teleportation protocol allowing the transfer of the
information associated to an unknown incoming state from one place
to another. For sake of comparison with our proposal we recall
briefly the standard teleportation protocol:\ Alice has an unknown
qubit $\left\vert \chi \right\rangle _{U}=\alpha \left\vert
0\right\rangle _{U}+\beta \left\vert 1\right\rangle _{U}$\ and she
wants to send it to Bob (in order to avoid confusion this general
description $\left\vert 0\right\rangle $ and $\left\vert
1\right\rangle $ are merely label indicating the qubit states,
without referring to charge occupation). They need to share a
maximally entangled two qubits state (e.g.$\left[ \left\vert
0_{A},1_{B}\right\rangle +\left\vert 1_{A},0_{B}\right\rangle
\right] $) where subscripts A and B denote respectively Alice and
Bob subspaces (normalization factors will be
ignored). The whole state can now be rewritten in terms of Bell basis ($%
\left\vert \phi ^{\pm }\right\rangle =$ $\left( \left\vert
0,0\right\rangle \pm \left\vert 1,1\right\rangle \right) ;$
$\left\vert \psi ^{\pm }\right\rangle =$ $\left( \left\vert
0,1\right\rangle \pm \left\vert 1,0\right\rangle \right) $):
\begin{multline}
\left\vert \Psi \right\rangle _{UAB}=\alpha \left\vert 0,0,1\right\rangle
_{UAB}+\alpha \left\vert 0,1,0\right\rangle _{UAB}+ \\
\beta \left\vert 1,0,1\right\rangle _{UAB}+\beta \left\vert
1,1,0\right\rangle _{UAB}= \\
\left\vert \phi ^{+}\right\rangle _{UA}\sigma _{x}\left\vert \alpha
\right\rangle _{B}-i\left\vert \phi ^{-}\right\rangle _{UA}\sigma
_{y}\left\vert \alpha \right\rangle _{B}+ \\
\left\vert \psi ^{+}\right\rangle _{UA}\left\vert \alpha \right\rangle
_{B}+\left\vert \phi ^{-}\right\rangle _{UA}\sigma _{z}\left\vert \alpha
\right\rangle _{B}
\end{multline}
having indicated with $\sigma _{i}$\ standard Pauli matrices.
Making a projective measure on $U$ and $A$ (Alice's qubits) and
transmitting via a classical channel the result, we are able to
recover the original unknown state in the Bob site without need of
measurements in the destination place. The Bell measurement
process can be performed (as suggested by G. Brassard and
coworkers) in two sequential steps \cite{brassard}: first, Bell
states
are rotated in the computational basis ($\left\vert 0,0\right\rangle $,$%
\left\vert 0,1\right\rangle $,$\left\vert 1,0\right\rangle
$,$\left\vert 1,1\right\rangle $), then the projective measure is
performed in this latter basis.

Here we propose a slightly modified procedure wherein the Bell
states involved are two instead of four; furthermore, we exploit
temporal evolution to perform the first step of Brassard method,
making simple the final one.

Our protocol exploits an adiabatic switching on of Coulomb
interaction between the qubit we want to teleport and the
entangled state. Now we deal with a system composed by three DQDs
(see figure 2), one of them is used to encode the unknown qubit
and the other two as entangled support. The Hamiltonian is
\begin{align}
H_{123456}& =-w\left( a_{3}^{\dag }a_{4}+a_{4}^{\dag }a_{3}\right)
-w^{\prime }(t)\left( a_{5}^{\dag }a_{6}+a_{6}^{\dag }a_{5}\right) +  \notag
\\
& -w^{\prime \prime }(t)\left( a_{1}^{\dag }a_{2}+a_{2}^{\dag }a_{1}\right)
+U\left( t\right) \left( n_{3}n_{6}+n_{4}n_{5}\right) +  \notag \\
& +U^{\prime }\left( t\right) \left( n_{1}n_{4}+n_{2}n_{3}\right)
\end{align}

where $U^{\prime }\left( 0\right) $and $w^{\prime \prime }(0)$ vanishe,
while $w^{\prime }(0)=w$.

Making use of encoding technique and entanglement generation above
described, the incoming overall state is
\begin{align}
\left\vert \Psi \right\rangle _{123456}& =\frac{1}{\sqrt{2}}\left( \alpha
\left\vert 0_{1},1_{2}\right\rangle +\beta \left\vert
1_{1},0_{2}\right\rangle \right) \times  \notag \\
& \left( \left\vert 0_{3},1_{4},0_{5},1_{6}\right\rangle +\left\vert
1_{3},0_{4},1_{5},0_{6}\right\rangle \right)
\end{align}
If $U^{\prime }\left( t\right) $ is adiabatically increased,\ the state
evolves and reaches its new ground state
\begin{align}
\left\vert \Psi \left( t\right) \right\rangle _{123456}& =\frac{1}{\sqrt{2}}%
(\alpha \left\vert 0_{1},1_{2},0_{3},1_{4},0_{5},1_{6}\right\rangle +  \notag
\\
& \beta \left\vert 1_{1},0_{2},1_{3},0_{4},1_{5},0_{6}\right\rangle )
\label{n123456}
\end{align}
a GHZ-like three qubits nonmaximally entangled state \cite{GHZ}.

So far we have described the coupling between unknown qubit and
entangled state. Next step represents the analogous of Bell
measurement. To prepare it we need to detach Bob QDs (5 and 6)
from the others and to start a temporal evolution of the state
which involves dots from $1$ to $4$. By instantaneously turning on
the tunneling $w^{\prime \prime }(t)$, and
turning off the tunneling $w^{\prime }(t)$ and the Coulomb interaction $%
U\left( t\right) $ (from now on the time will be measured starting
form the switching instant), the system is forced to belong to a
state in which dots from $1\ $to $4$ evolve following the
Hamiltonian of Eqn. \ref{Hamiltonian1} (with appropriate indices),
while Bob's dots are frozen. Neglecting terms of the order of
$w/U$, $\left| 0_{1},1_{2},0_{3},1_{4}\right\rangle $\ \ evolves
in $\left( \cos \omega t\left|
0_{1},1_{2},0_{3},1_{4}\right\rangle
+i\sin \omega t\left| 1_{1},0_{2},1_{3},0_{4}\right\rangle \right) $ and $%
\left| 1_{1},0_{2},1_{3},0_{4}\right\rangle $ \ evolves in $\left(
\cos \omega t\left| 1_{1},0_{2},1_{3},0_{4}\right\rangle +i\sin
\omega t\left| 0_{1},1_{2},0_{3},1_{4}\right\rangle \right) $
where $\omega =2w^{2}/U$. Thus, the whole state becomes
\begin{align}
\left\vert \Psi\left( t\right) \right\rangle _{123456} & =\frac{1}{\sqrt {2}}%
(\left\vert 0_{1},1_{2},0_{3},1_{4}\right\rangle \left\vert \chi
^{+}\right\rangle _{56}+  \notag \\
& +i\left\vert 1_{1},0_{2},1_{3},0_{4}\right\rangle \left\vert
\chi ^{-}\right\rangle _{56})
\end{align}
having introduced $\left| \chi ^{\pm }\left( t\right) \right\rangle _{56}=%
\left[ \left( \cos \omega t\right) \alpha \left|
0_{5},1_{6}\right\rangle
\pm i\left( \sin \omega t\right) \beta \left| 1_{5},0_{6}\right\rangle %
\right] $.Waiting a suitable time ($\omega t=\pi /4$) we obtain,
associated with two orthogonal computational states on the four
Alice's dots, $\alpha \left| 0_{5},1_{6}\right\rangle +i\beta
\left| 1_{5},0_{6}\right\rangle $ and $\alpha \left|
0_{5},1_{6}\right\rangle -i\beta \left| 1_{5},0_{6}\right\rangle
$. Measuring the charge on a dot (e.g. the number 1), Alice
transmits the result as classical bit to Bob, that can choose the
correct unitary rotation to perform in order to completely recover
$\left| \chi \right\rangle $ on its site. Note that due to the
nonlinearity of interactions involved in this model, there are no
conceptual obstacles for which Bell measurements cannot reach a
100\% of success probability \cite {diporto}.

\begin{figure}
  \includegraphics[width=2.8928in]{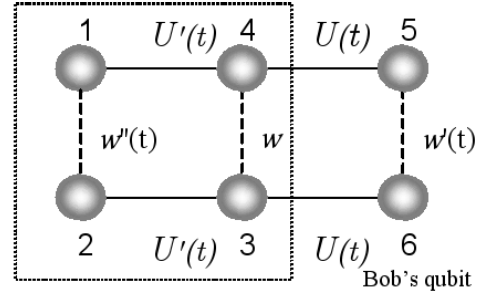}
  \caption{Final step of quantum
teleportation: Bob's qubit is separated by others QD's which
evolve providing a Bell measurement process}\label{fig2}
\end{figure}

The teleportation protocol described allows an information
transfer over typical interdot distances, but it's straightforward
to extent this model on a array of $N$ DQDs aligned and coupled
one-to-one by Coulomb interaction. In effect, the state of last
$N-1$ DQDs can be arranged, by means of adiabatic couplings
between them, in a maximally entangled state
\begin{equation}
\frac{1}{\sqrt{2}}\left[ \left\vert 0,1,0,1,0,1,.....,0,1\right\rangle
+\left\vert 1,0,1,0,1,0,.....,1,0\right\rangle \right]
\end{equation}
The coupling with the first, adjunctive, DQD used to encode a
qubit, like in the foregoing discussion, will drive the system in
the generalization of Eqn. \ref{n123456}, allowing the final step
of teleportation. Thus, we deal with a quantum channel with high
fidelity: losses are only due to contributes to ground state
coming from the time evolution such as in Eqn. \ref{tinfinito},
and can be strongly reduced with a suitable choice of coupling
parameters.

\section{Conclusions}

To summarize, we have proposed a solid state implementation of a
quantum teleportation protocol. Qubits are represented by DQDs
coupled by tunneling with one excess electron staying coherently
in the first or in the second dot. Coherent oscillations in these
structures, experimentally observed, show how it's possible to
conceive a network of DQDs. In this paper we have illustrated a
scheme to generate entanglement between adjacent qubits using
adiabatic changes in the system parameters. Extending this
technique also to nonadiabatic variations we have formulated a
proposal for optimal quantum teleportation. Furthermore, this
technique can be generalized to an array of $N$ DQDs embedded in a
chain, allowing the transmission with high fidelity over
relatively long distances of the unknown qubit.

\section{Acknowledgments}

One of us (F.d.P.) acknowledges the ospitality of LEPES and important
discussions with D. Feinberg and B. Chakraverty. Enlightening criticism of
P. Di Porto is also acknowledged.


\begin{thebibliography}{99}
\bibitem{bendiv}  C. H. Bennett and D. P. DiVincenzo, \textrm{Nature}
(London) \textbf{404}, 247 (2000).

\bibitem{tele}  C. Bennett, G. Brassard, C. Crepeau, R. Jozsa, A. Peres, and
W. Wootters, \textrm{Phys. Rev. Lett.} \textbf{70}, 1895 (1993).

\bibitem{optics}  B. Bouwmeester, J.W. Pan, K. Mattle, M. Eibl, H.
Weinfurter, and A. Zeilinger, \textrm{Nature} (London) \textbf{390}, 575
(1997); D. Boschi, S. Branca, F. De Martini, L. Hardy, and S. Popescu,
\textrm{Phys. Rev. Lett.} \textbf{80}, 1121 (1998); A. Furusawa, J. L. S\o
rensen, S. L. Braunstein, C. A. Fuchs, H. J. Kimble, and E. S. Polzik,
\textrm{Science} \textbf{282}, 706 (1998).\textbf{\ }

\bibitem{NMR}  M. A. Nielsen, E. Knill and R. Laflamme, \textrm{Nature}
(London) \textbf{396}, 52 (1998).

\bibitem{feinberg}  O. Sauret, D. Feinberg and Th. Martin, \textrm{Eur.
Phys. J. B} \textbf{32}, 545 (2003).

\bibitem{reina}  J. H. Reina and N. F. Johnson, \textrm{Phys. Rev. A}
\textbf{63}, 012303 (2001).

\bibitem{zanardi}  P. Solinas, P. Zanardi, N. Zangh\`{\i}, and F. Rossi,
\textrm{Phys. Rev. A} \textbf{67}, 052309 (2003).

\bibitem{lombardi}  E. Lombardi, F. Sciarrino, S. Popescu, and F. De
Martini, \textrm{Phys. Rev. Lett.} \textbf{88}, 070402 (2002); Hai-Wong Lee
and J. Kim, \textrm{Phys. Rev. A} \textbf{63}, 012305 (2001).

\bibitem{mompart}  J. Mompart, K. Eckert, W. Ertmer, G. Birkl, and M.
Lewenstein, \textrm{Phys. Rev. Lett.} \textbf{90}, 147901 (2003).

\bibitem{fujisawa}  T. Hayashi, T. Fujisawa, H. D. Cheong, Y. H. Jeong, and
Y. Hirayama, \textrm{Phys. Rev. Lett.} \textbf{91}, 226804 (2003).

\bibitem{brandes}  F. Renzoni and T. Brandes, \textrm{Phys. Rev. B} \
\textbf{64, }245301 (2001).

\bibitem{loss}  D. Loss and D. P. DiVincenzo, \textrm{Phys. Rev. A} \textbf{%
57}, 120 (1998).

\bibitem{messiah}  A. Messiah, \textit{Quantum Mechanics, Vol II }%
(North-Holand, Amsterdam, 1961).

\bibitem{steffen}  M. Steffen, W. van Dam, T. Hogg, G. Breyta, and I.
Chuang, \textrm{Phys. Rev. Lett.}\textbf{\ 90}, 067903 (2003).

\bibitem{brassard}  G. Brassard, S. Braunstein and R. Cleve, \textrm{Physica
D}, \textbf{120}, 43 (1998).

\bibitem{GHZ}  D. M. Greenberger, M. Horne, A. Shimony, and A. Zeilinger,
\textrm{Am. J. Phys.} \textbf{58}, 1131 (1990).

\bibitem{diporto}  E. DelRe, B. Crosignani, and P. Di Porto, \textrm{Phys.
Rev. Lett.} \textbf{84}, 2989 (2000); L. Vaidman and N. Yoran, \textrm{Phys.
Rev. A} \textbf{59}, 116 (1999); N. Lutkenhaus, J. Calsamiglia, and K.-A.
Suominen, \textrm{Phys. Rev. A} \textbf{59}, 3295 (1999).
\end{thebibliography}
\end{document}